\documentclass[twocolumn,showpacs,superscriptaddress,amsmath,amssymb]{revtex4}
\usepackage{graphicx}
\usepackage{dcolumn}
\usepackage{bm}

\def\e{{\rm e}}  \def\d{{\rm d}}
       
   \def\ta{\theta}  
    
\def\ie{{\it i.e. }}   \def\cf{{\it cf. }}

\def\bcc{\begin{center}} \def\ecc{\end{center}}
\def\beq{\begin{equation}} \def\eeq{\end{equation}}
\def\bea{\begin{eqnarray}} \def\eea{\end{eqnarray}}
\def\beqa{\begin{eqnarray}}  \def\eeqa{\end{eqnarray}}
  
\def\btbl{\begin{tabular}} \def\etbl{\end{tabular}}
\def\bpm{\begin{pmatrix}} \def\epm{\end{pmatrix}}
\def\btbb{\begin{tabbing}} \def\etbb{\end{tabbing}}
\def\btm{\begin{itemize}} \def\etm{\end{itemize}}
  \def\hs{\hskip}
  \def\nbr{\nonumber}
    
\def\ched{\end{CJK*} \end{document}}

\def\Iarrow{\longrightarrow \hs-19pt^{\rm (I)}\hs9.5pt}

\begin{document}

\title{A Monte Carlo Study on the Identification of Quark-gluon Fusion Product \\
in QCD-instanton Induced Processes in Deep-inelastic Scattering}
\author{Xu Mingmei and Liu Lianshou\footnote{Email: liuls@iopp.ccnu.edu.cn}}
\affiliation{Institute of Particle Physics, Huazhong Normal
University, Wuhan 430079, China}

\begin{abstract}
Different methods to reconstruct the quark-gluon fusion product
and current jet are tried in deep-inelastic e-p scattering events
with instanton as background generated by QCDINS Monte Carlo code.
A comparison of these methods are performed and a good method is
found which can reconstruct well the energies of current jet and
instanton product as well as the mass of the latter. The isotropy
property of the instanton product and jet are calculated and
compared. A parameter characterizing the degree of ``hardness'' of
the instanton product is presented.
\end{abstract}


\maketitle
\section{Introduction}

In the Standard Model, both the strong and the electro-weak
interactions are described by non-Abelian gauge theory. In these
theories, the ground state has a rich topological structure,
associated with non-perturbative fluctuations of the gauge field,
called instantons~\cite{belavin,hooft1,hooft2,hooft3}, which
represent tunneling transition between topologically non-equivalent
vacua. In the strong sector, described by QCD, instantons are
non-perturbative  fluctuations of the gluon field. Deep-inelastic
scattering (DIS) offers a unique opportunity to discover the
processes induced by QCD instantons. The rate is calculable within
``instanton-perturbative theory" and is found to be
sizable~\cite{Moch,Ringwald,Ringwald2}.

QCD-instanton induced process leads to a characteristic final
state, which allows instanton induced events to be distinguished
from the normal DIS processes. For a leading graph in
QCD-instanton induced e-p collision, \cf Fig. 1, the incident
lepton emits a photon, with 4-momentum $q$, which in turn
transforms into a quark-antiquark pair. One of these quarks with
4-momentum $q''$ hadronizes to form the {\it current jet}. The
other quark, with 4-momentum $q'$, fuses with a gluon (4-momentum
$g$) from the proton in the presence of an instanton. The $q'g$
interaction is called the hard subprocess of photon-gluon-fusion
process. For simplicity, we will in the following call the hadron
system produced from the fusion of quark-gluon in the presence of
an instanton as {\it instanton final state} (IFS).

\begin{figure}
\centering
\includegraphics[width=1.7in]{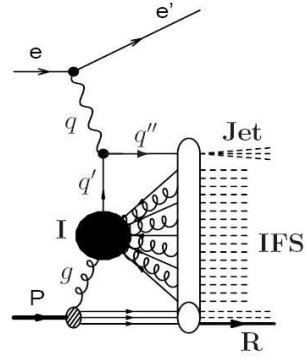}
\caption{\ \ The leading QCD-instanton induced process in the DIS
regime of e-p scattering.}
\end{figure}

The phenomenological characteristics of instanton induced events
can be summarized as follows~\cite{phen}:

\begin{itemize}

\item In the hard subprocess exactly one $q\overline{q}$ pair of
each of the $n_{f}$ kinematically accessible quark flavors
participates in the hard subprocess of each event, either as
incoming or as outgoing object. This feature is often termed as
``flavour democracy". The number of gluons emitted per event
follows a Poisson distribution with average value $\langle
n_{g}\rangle\sim O(1/\alpha_{s})\sim 3$. The $n_{g}$ gluons plus
$2n_{f}-1$ outgoing (anti-)quarks give rise to a high multiplicity
final state.

\item The particles produced from the quark-gluon fusion in the
presence of instanton, \ie the IFS, are expected to be isotropically
distributed in their center of mass frame.

\item From theoretical point of view, the most prominent
characteristic of instanton induced events, which is also the most
difficult feature to be observed experimentally, is the {\it
chirality violation}. In instanton induced events, all quarks
produced in the hard subprocess are emitted with the same
handedness, that means
\begin{equation}
\gamma^{*}+g\Iarrow \sum_{n_{f}}(q_{R}+\overline{q}_{R})+n_{g}g,
\quad (I\rightarrow \overline{I},R\rightarrow L).
\end{equation}

\end{itemize}

Although the existence of instanton is required by the Standard
Model, the experimental evidence is still lacking. HERA offers a
unique chance to discover instanton, which would be a confirmation
of an essentially non-perturbative Standard Model prediction.

Two experiments, H1 and ZEUS, have reported their search for
QCD-instanton induced processes in DIS at HERA~\cite{H1,ZEUS}. While
an excess of events with instanton-like topology has been observed,
it cannot be claimed significantly given the uncertainty of
simulation and theory. Upper limits on the cross-section for
instanton induced processes are set dependent on the kinematic
domain considered.

One of the chief problems in instanton-searching is to identify
the IFS, current jet and proton remnant. Only when the IFS and jet
are identified precisely the reconstruction of the kinematic
variables such as $Q'^{2}\equiv-q'^{2}$, $x'\equiv Q'^{2}/(2g\cdot
q')$ can be good and the discriminating variables for
instanton-searching can be reliable.

At partonic level the separation of different parts is clear, \cf
Fig. 1. The mother parton of current jet is the quark $q''$. The
``mother'' of IFS is the quark $q'$ and gluon $g$ in the presence
of I. The best identification of IFS and jet would be to minimize
the differences between the energies and masses of IFS and current
jet with those of their ``mother''
------ $q'+g$ and $q''$.
However, this could be achieved only in Monte Carlo study but not
in real experiment, because the 4-momenta at partonic level are
unmeasurable.

Monte Carlo simulation has been applied to the problem in
consideration by some
authors\;~\cite{H1,Gerigk,Koblitz,Carli,ZEUS,Sievers,Hillert}.
Their main goal was concentrated on finding some methods that can
be used in experiments, and the information at partonic level has
been used only for evaluating the methods.

For jet finding, both the cone
algorithm~\cite{H1,Gerigk,Koblitz,Carli} and the $k_{t}$-cluster
algorithm~\cite{ZEUS,Sievers,Hillert} have been used. The cone
size has been chosen to optimize the reconstruction of
$Q'^{2}$~\cite{Gerigk}. The IFS is identified by a pseudo-rapidity
band with width 1.1, where the width is chosen for the expectation
of an isotropic IFS in its rest frame. While the $k_{t}$ algorithm
has not done any optimization and the identification of IFS was
through quadrant~\cite{Sievers} according to its proper position.
Although these methods play an important role in the present
searching strategy, the optimization needs improving.

The aim of the present paper is to re-consider the Monte Carlo
study of the problem and identify IFS through directly optimizing
its energy and mass as well as the energy of current jet. The
resulting IFS is of high precision and can be used in the
theoretical study of instanton physics. The method used here is
not directly applicable to real experiments but can provide
instructions for the IFS identification in experimental study.

\def\bll{\bf\small}

The layout of the paper is as follows. In Sec. II some information
at the partonic level of the instanton induced events will be
presented. Several new methods for the reconstruction of IFS in
Monte Carlo simulation will be proposed and compared in Sec. III.
A good method is found which can reconstruct well the energies of
current jet and instanton as well as the mass of the latter. The
isotropy property of the obtained IFS will be discussed in Sec. IV
in comparison with that of the current jet. Sec. V is a summary.

\section{Basic distributions with partonic information}
Our study is based on the Monte Carlo generator
QCDINS~\cite{qcdins,Gibbs} for instanton induced events. It is a MC
package to simulate quark-gluon fusion process $q'+g\Iarrow X$ in
the background of instanton. It acts as a hard subprocess generator
embedded in the HERWIG~\cite{herwig1,herwig2} program. The
subsequent fragmentation and hadronization are handled by HERWIG. In
the following the default parameters of the QCDINS 2.0 version are
used, i.e. $x'>0.35$, $Q'^{2}>113$ ${\rm GeV}^{2}$ and the number of
flavors is set to be $n_{f}=3$.

Final states in instanton induced DIS consist of 4 subsystems: 1-
scattered lepton, 2- hadrons from current quark, 3- hadrons
created from the fusion of quark and gluon in the background of
instanton ({\it instanton final state} for short), 4- proton
remnant group of particles, \cf Fig.\;1. The reconstruction is
based on the assumption that the color forces among the current
jet, the instanton part and the proton remnant are weak so that
each part can be approximately separated and the momentum of
hadronic final state of each part is approximately equal to those
of the corresponding part at partonic level.

In HERWIG the scattered electron is easily identified by particle
ID and status code. We discard it first. The particles left are
the mixture of current jet (C for short), instanton final state (I
for short) and proton remnant (R for short).


After rejecting the scattered lepton, all the objects in the
hadronic final state are boosted to the hadronic center-of-mass
frame (hcm). This frame is defined by ${\bf q}+{\bf P}=0$, where
${\bf q}$ and ${\bf P}$ are the 3-momenta of the exchanged photon
and proton, respectively. The positive $Z$-direction of hcm frame is
defined in the direction of photon momentum. In the hcm frame, in
order to see the jet signal, we remove the $\phi$ angle of current
quark for every final state particles in every event. That means
letting the $\phi$ angle of current quark equal 0 and hence the jet
fragments are around $\phi=0$. The 2D distributions of $\theta$ and
$\phi$ for all final state particles are plotted in Fig.2. The 1D
distribution of $\theta$ for all the final state particles is
plotted in Fig.3\;(a). Also plotted in Fig.3\;(b)-(d) is the
partonic information of C, I and R. The angle $\theta$ is measured
against the direction of photon.

\begin{figure}
\centering
\includegraphics[width=3.in]{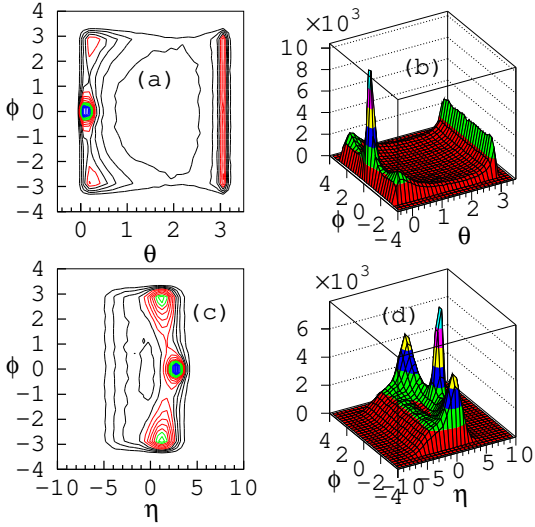}
\caption{\ \ (Color online) 2D plot in hcm after removing the
$\phi$ angle of current quark. (a)(b) --- $\phi$ vs. $\theta$ in
contour and surface plots, respectively; (c)(d) --- $\phi$ vs.
$\eta$ ($\eta$ is defined as $-\log(\tan\frac{\theta}{2})$) in
contour and surface plots, respectively.}
\end{figure}
\begin{figure}
\centering
\includegraphics[width=3.in]{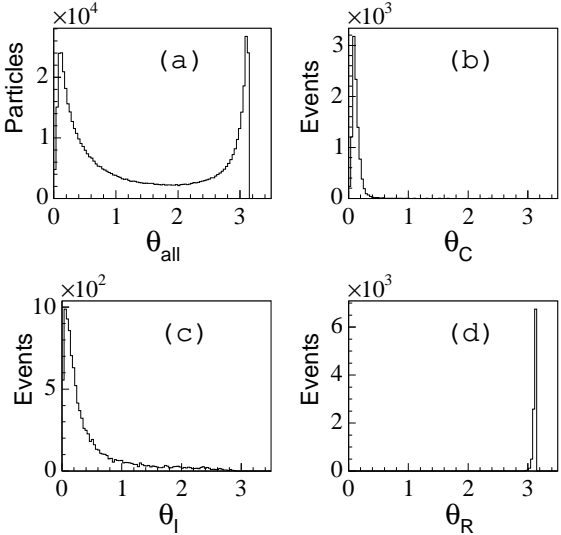}
\caption{\ \ The distribution of $\theta$ defined in hcm frame.
(a) for all final state particles; (b) (c) (d) are calculated from
the momenta of current quark, instanton part and proton remnant,
respectively, at partonic level.}
\end{figure}

\section{Three methods for IFS identification}
\def\ve{\varepsilon}
\begin{figure}
\centering
\includegraphics[width=3.in]{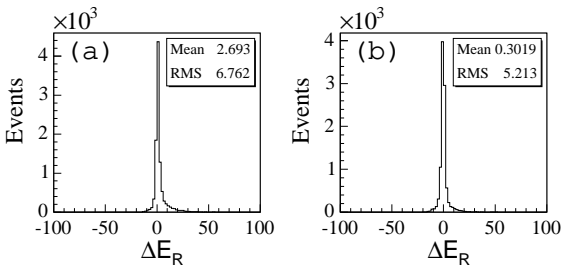}
\caption{\ \ The energy reconstruction errors of proton remnant by
doing two cuts in $\theta$ in hcm£¬(a) $\theta_{\rm
cut}=\frac{\pi}{2}$, (b) $\theta_{\rm cut}=\frac{2\pi}{3}$. The
legends show the Mean and RMS.}
\end{figure}

\noindent {\bf Method 1: $p_z$-sorting} \vskip3mm

Fig. 3(a) illustrates that the final state particles distribute
mainly around two back-to-back directions ------ $\theta=0$ and
$\pi$. Around $\theta=0$ are C and the main part of I, around
$\theta=\pi$ are R and a little I. The method to discard R in ZEUS
experiment~\cite{ZEUS} is to do a $\theta=\frac{\pi}{2}$ cut, taking
particles with $\theta^{\rm hcm}>\frac{\pi}{2}$ as proton remnant.
In fact, as can be seen from Fig. 3(a), the peak around $\theta=\pi$
is narrower than that around $\theta=0$. So, $\theta_{\rm cut}$
should be in a place on the right of $\frac{\pi}{2}$. We find that
$\theta_{\rm cut}=\frac{2\pi}{3}$ is better than $\theta_{\rm
cut}=\frac{\pi}{2}$, which results in a better reconstruction of the
energy of proton remnant, \cf Fig. 4.

We define the reconstruction error of a variable $Y$ as the
difference between the reconstructed value and the true value before
hadronization, {\it i.e.}
\begin{equation}
\Delta Y=\frac{Y_{\rm rec}-Y_{0}}{|Y_{0}|}\times 100\%,
\end{equation}
 where $Y_{\rm rec}$ represents the reconstructed
value, $Y_{0}$ represents the true value before hadronization. In
contrast to the case of $\theta_{\rm cut}=\frac{\pi}{2}$ (see
Fig.\;4(a)), the energy reconstruction error for R after
$\theta_{\rm cut}=\frac{2\pi}{3}$ (see Fig.\;4(b)) is focused at 0
and has a smaller RMS. In other words, the $\theta_{\rm
cut}=\frac{2\pi}{3}$ method reconstructs the energy of R more
precisely.

\begin{figure}
\centering
\includegraphics[width=3.in]{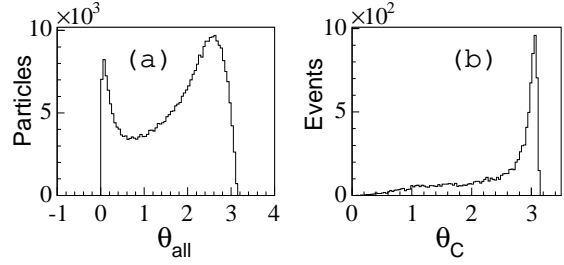}
\caption{\ \ $\theta$ distribution in cm2, (a) for final state
particles (C+I), (b) for instanton before hadronization.}
\end{figure}
\begin{figure}
\includegraphics[width=3.2in]{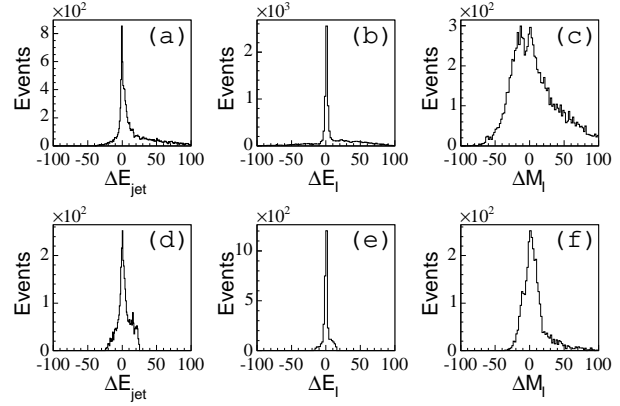}
\caption{\ \ The reconstruction errors for jet energy $\Delta
E_{\rm jet}$, instanton energy $\Delta E_{\rm I}$ and instanton
mass $\Delta M_{\rm I}$ in $p_{z}$-sorting method. Bottom panels
are the results of a cut on $\Delta E<10\%$ of top panels.}
\end{figure}
\begin{figure}
\centering
\includegraphics[width=2.in]{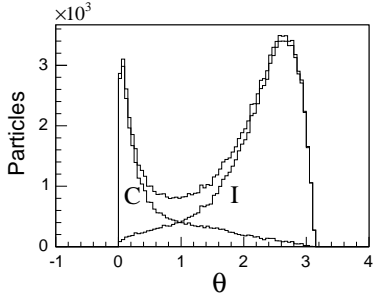}
\caption{\ \ The $\theta$ distribution in cm2 of identified C and
I by $p_z$-sorting method.}
\end{figure}
After discarding R by doing $\theta_{\rm cut}=\frac{2\pi}{3}$, the
left particles are the mixture of C and I. Boost (C+I) to their
c.m.s. frame (cm2 for short). Take the momentum direction of
current quark, {\it i.e.} the direction of the jet axis, as the
$Z$ axis of cm2. The $\theta$ distribution of final state
particles (C+I) in cm2 is shown in Fig.\;5(a) and that of
instanton before hadronization in Fig.\;5(b). The two peaks in
Fig.\;5(a) represent current jet and IFS respectively. Since the
$Z$ axis is in the direction of jet axis, the peak around
$\theta=0$ represents the current jet particles, while the other
peak, around $\theta=\pi$, represents the IFS.

Renumber these $n$ particles by their $p_{z}$, sorting them
according to $p_{z1}>p_{z2}>\cdots
>p_{zn}$. Accumulating the energy from particle 1 to particle $k$ gives
$E_{k}=\ve_{1}+\ve_{2}+...+\ve_{k}$ ($\ve_{i}$ represents the
energy of the $i$-th particle); simultaneously accumulating the
energy from particle $k+1$ to $n$ gives
$E'_{k}=\ve_{k+1}+\ve_{k+2}+...+\ve_{n}$. Taking particles from 1
to $k$ as current jet, from $k+1$ to $n$ as IFS, the energy
reconstruction errors for jet and instanton are $\Delta E_{\rm
jet}=\frac{E_{k}-E_{\rm C}}{E_{\rm C}}\times 100\%$ and $\Delta
E_{\rm I}=\frac{E'_{k}-E_{\rm I}}{E_{\rm I}}\times 100\%$,
respectively, where $E_{\rm C}$ is the energy of current parton,
$E_{\rm I}$ is the energy of the quark and gluon included in IFS
at partonic level. The value of the parameter $k$ is chosen to
make $\Delta E=0.4\times |\Delta E_{\rm jet}|+0.6\times |\Delta
E_{\rm I}|$ minimum.

The reconstruction errors for the energy of jet, the energy of
instanton and the mass of instanton are shown in Fig's.\;6\;(a),
(b) and (c), respectively. It can be seen that large error events
occur with some probability and the mass of instanton has not been
reconstructed well. The criterion on which we judge for a good
reconstruction quality is based on that the reconstruction error
centralizes around 0 and has a narrow RMS. After doing a cut
$\Delta E<10\%$, both are improved ------ large error events have
been cut out and the mass of instanton is reconstructed better at
the expense of throwing away 67\% of the events, \cf
Fig's.\;6\;(d), (e), (f).

Up to now, the current jet and IFS are separeated. The $\theta$
distribution of each part in cm2 are shown in Fig.\;7. Jet
particles centralize around $\theta=0$, IFS centralizes around
$\theta=\pi$. At intermediate $\theta$ the two parts overlap. They
are separated by their $p_{z}$.

\vskip3mm \noindent {\bf Method 2: 2D-cut} \vskip3mm

Based on the 2D plot shown in Fig.\;2 and enlightened by the
quadrant method (or quadrate-cut method) used in~\cite{Sievers},
an improved polygonal cut is used in $\eta$-$\phi$-plane to
isolate each part. Considering that the relative position of C and
I in $\eta$-$\phi$-plane strongly depends on the virtuality
$Q^{2}$ of photon, the polygonal cut is running with $Q^{2}$, as
shown in Fig.\;8. Using these cuts each part is identified. The
reconstruction errors are shown in the upper panels of Fig.\;10.
\begin{figure}
\centering
\includegraphics[width=3.5in]{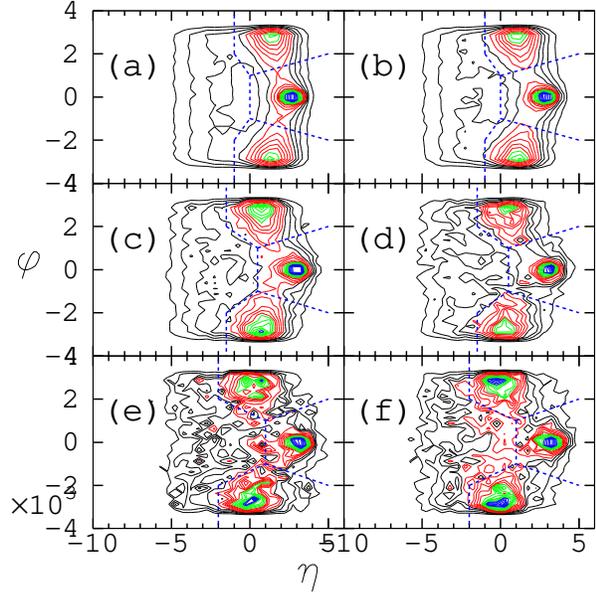}
\caption{\ \ (Color online) Running of the polygonal cuts (dashed
lines) with $Q^{2}$. The intervals of $Q^{2}$ for (a), (b), (c),
(d), (e), (f) are $[113, 200]$, $[200, 400]$, $[400, 600]$, $[600,
800]$, $[800, 1000]$, $[1000, \infty]\; {\rm GeV}^{2}$,
respectively.}
\end{figure}

\vskip3mm\noindent {\bf Method 3: $r$-sorting in 2D plane}\vskip3mm
\begin{figure}
\centering
\includegraphics[width=3.in]{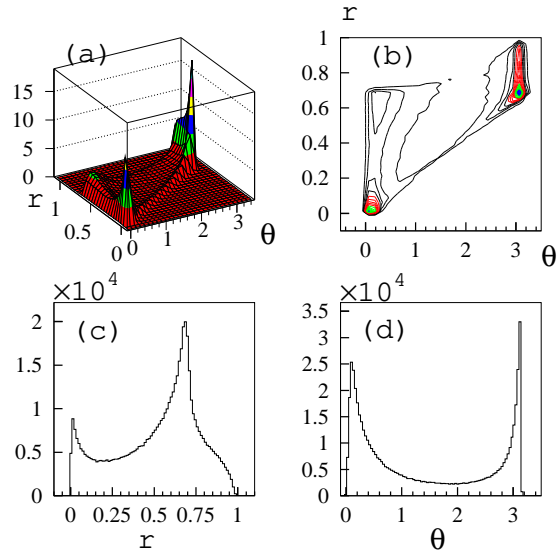}
\caption{\ \ (Color online) (a) (b) are 2D plots $r$ vs. $\theta$
in hcm, surface and contour plots, respectively; (c) (d) are 1D
distribution for $r$ and $\theta$ respectively. All the variables
are calculated in hcm.}
\end{figure}

Comparing the two methods mentioned above it can be seen that the
$p_{z}$-sorting method gives better reconstruction. However, it
disregards totally the information contained in the
$\theta$-$\phi$-plane, \cf Fig.\;2, which has been taken into
account in the 2D-cut method. Therefore, we try to combine the two
methods.

Let us define a distance in  $\theta$-$\phi$-plane,
\begin{equation}
r(\ta,\phi)=\sqrt{\frac{(\frac{\theta-\theta_{0}}{\pi})^{2}+(\frac{\phi-0}{\pi})^{2}}{2}},
\end{equation}
where $(\theta_{0},\phi=0)$ is the position of current quark. This
variable measures how far every final state particle is from the
jet axis. Choosing an appropriate value for $r_0$ the particles
with $r(\ta,\phi)<r_{0}$ are attributed to jet. The distribution
of $r(\ta,\phi)$ is plotted in Fig.\;9.

First discard R by 1D cut $\theta^{\rm hcm}>\frac{2\pi}{3}$.
Renumber the left $n$ particles by their $r$, let
$r_{1}<r_{2}<\cdots <r_{n}$. Accumulating the energy from particle 1
to particle $k$ gives $E_{k}=\ve_{1}+\ve_{2}+...+\ve_{k}$,
simultaneously accumulating energy from $k+1$ to $n$ gives
$E'_{k}=\ve_{k+1}+\ve_{k+2}+...+\ve_{n}$. Taking particles from 1 to
$k$ as current jet, from $k+1$ to $n$ as IFS. The value of $r_{0}$
is chosen to optimize the energy reconstruction, \ie make $\Delta
E=0.4\times |\Delta E_{\rm jet}|+0.6\times |\Delta E_{\rm I}|$ the
minimum. The reconstruction errors are shown in Fig's.\;10\;(d), (e)
and (f).
\begin{figure}
\includegraphics[width=3.2in]{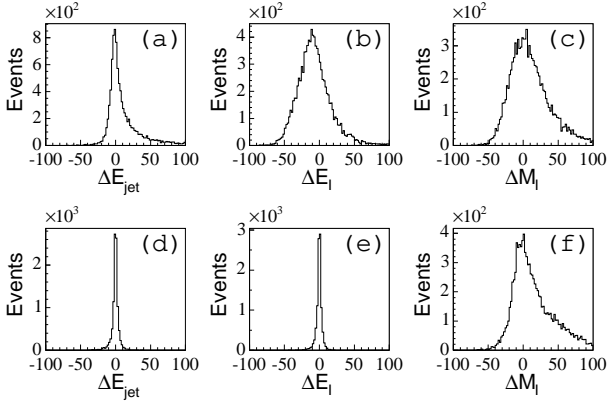}
\caption{\ \ The reconstruction errors for jet energy $\Delta
E_{\rm jet}$, instanton energy $\Delta E_{\rm I}$ and instanton
mass $\Delta M_{\rm I}$ in 2D-cut method (upper panels). Shown in
the bottom panels are the results of $r$-sorting method.}
\end{figure}

It is obvious that $r$-sorting in 2D plane gives the best
reconstruction among the 3 methods described above.

\section{A comparison of the isotropy degree of IFS and jet}

IFS are expected to be isotropically distributed in its rest
frame. Jet fragments are anisotropy, being axially symmetric. To
quantify the degree of isotropy we use the sphericity.

Sphericity is a measure of how isotropically a collection of
4-momenta is distributed in three dimensions. A normalized
momentum tensor is calculated from the momenta of the particles in
consideration, \ie
\begin{equation}
S^{\alpha,\beta}=\frac{\sum_{i}P_{i}^{\alpha}P_{i}^{\beta}}{\sum_{i}|P_{i}|^{2}},
\end{equation}
where $\alpha$, $\beta$=1, 2, 3 correspond to the $x$, $y$ and $z$
components. By standard diagonalization of $S^{\alpha,\beta}$ one
find three eigenvalues
$\lambda_{1}\geq\lambda_{2}\geq\lambda_{3}$, with
$\lambda_{1}+\lambda_{2}+\lambda_{3}=1$. The sphericity of the
event is then defined as
\begin{equation}
S=\frac{3}{2}(\lambda_{2}+\lambda_{3}).
\end{equation}
An ideally isotropic system with infinite multiplicity has $S=1$,
while a perfectly anisotropic system has $S=0$.

For a comparison, the sphericity of IFS and current jet in their
own rest frames identified by the above-mentioned $r$-sorting
method are calculated and the corresponding sphericity
distributions are shown in Fig's.\;11\;(a) and (b), respectively.
The mean values are \beq \bar S_{\rm IFS}=0.53\pm 0.18 , \qquad
\bar S_{\rm jet}=0.20\pm 0.17. \eeq

Since the value of sphericity depends on multiplicity, we
constructed an ideal sample, which has the same multiplicity as
the IFS sample but has ideally isotropic momentum distribution for
every event, \ie using the thermalized momentum distribution,
Eq.\;(7) below, for each component. The resulting sphericity is
drawn in Fig's.\;11\;(a) as dashed line. The corresponding Mean is
$\bar S^{\rm idea}=0.78\pm 0.09$. The mean sphericity for IFS
nearly equal to the ideally isotropic value means that IFS is
approximately isotropic in its rest frame. The jet is destined to
get low value of sphericity for its axial symmetry.

In view of the approximate isotropy of IFS we plot the
distribution of the three momentum components $p_x, p_y, p_z$ of
IFS in its c.m. frame as shown in Fig.\;12\;(a). They coincide
approximately. The distribution of the average $\bar p_i$ of these
three components is shown in Fig.\;12\;(b).

For a thermalized system the momentum component distribution is \beq
\d{\rm P}(p_i)=\frac{1}{\sqrt{2\pi mk_BT}}\e^{-p_i^2/(2mk_{B}T)}\d
p_i, \quad i=x,y,z.\eeq Fitting the $\bar p_i$ distribution shown in
Fig.\;12\;(b) to this formula we get the ``temperature'' of IFS as
\bea T&=& 863.86\;{\rm MeV\ \  for\ \  } m=m^\pi, \nbr \\
&=& 244.22\;{\rm MeV\ \  for\ \  } m=m^{\rm K}, \nbr \\
&=& 128.50\;{\rm MeV\ \ for\ \  } m=m^{\rm p}.\eea This
``temperature'' is actually not a thermal one, since at such high
temperature there will be deconfinement and no hadron can survive.

The ``temperature'' $T$ in Eq's.\;(8) is a parameter
characterizing the phase-space distribution of particles in IFS.
Its high value means that the quark-gluon fusion in the presence
of instanton is a very hard process.

\begin{figure}
\centering
\includegraphics[width=3.5in]{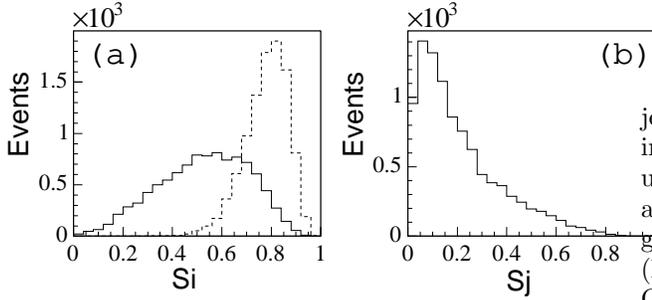}
\caption{\ \ Sphericity for IFS (a) and current jet (b) in their
respective rest frame. Dashed line in (a) shows the sphericity for
a ideally isotropic model with the same multiplicity as the IFS
sample.}
\end{figure}
\begin{figure}
\centering
\includegraphics[width=3.5in]{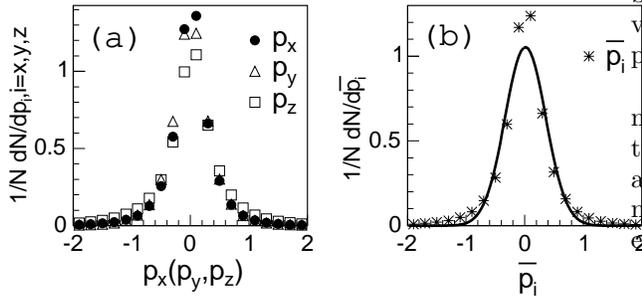}
\caption{\ \ (a) Momentum-component distribution of IFS in its
rest frame; (b) the distribution of the mean value of the three
components, fitted to thermalized momentum distribution.}
\end{figure}

\section{Summary}

The identification of instanton final state, current jet and proton
remnant in instanton induced deep inelastic scattering is studied
using Monte Carlo simulation. Different methods  ------
$p_{z}$-sorting, 2D-cut and $r$-sorting, for the reconstruction of
the quark-gluon fusion product in the background of instanton (IFS)
and current jet are tried and compared using QCDINS Monte Carlo
event generator. A method to optimize the energy reconstruction is
applied, which can reconstruct well the energies of current jet and
instanton as well as the mass of the latter. It is found that
$r$-sorting method gives the best reconstruction.

The sphericities of IFS and current jet identified by the
$r$-sorting method are calculated. The high value of sphericity
for IFS means that IFS is approximately isotropic in its rest
frame.

The momentum distribution of IFS approximately mimics a
thermalized distribution. The ``temperature'' fitted is rather
high, which can be taken as a charateristic parameter for
measuring the ``hardness'' of the quark gluon fusion process in
the presence of instanton.

 \hskip 10mm

{\bf Acknowledgement} \ This work is supported by National Natural
Science Foundation of China under project 10475030 and 10375025 as
well as the Cultivation Fund of the Key Scientific and Technical
Innovation Project£¬Ministry of Education of China NO
CFKSTIP-704035. One of the authors, Xu Mingmei, thanks
B.~B.~Levchenko for enlightening discussion.

\end{document}